\documentclass[aps,prb,longbibliography,floatfix,reprint]{revtex4-1}
\usepackage{amsmath}
\usepackage{longtable}
\usepackage{amsfonts}
\usepackage{graphicx}
\usepackage{color}
\usepackage{soul,xcolor}
\setstcolor{red}   

\begin{document}

\title{Temperature in a Peierls-Boltzmann Treatment of Nonlocal Phonon Heat Transport}

\author{Philip B. Allen}
\email{philip.allen@stonybrook.edu}
\affiliation{Department of Physics and Astronomy, Stony Brook University, Stony Brook, NY 11794-3800, USA}

\author{Vasili Perebeinos}
\affiliation{Skolkovo Institute of Science and Technology, 3 Nobel Street, Skolkovo, Moscow Region 143025, Russia}

\date{\today}


\begin{abstract}
In nonmagnetic insulators, phonons are the carriers of heat.  If heat enters in a region and temperature is measured
at a point within phonon mean free paths of the heated region, ballistic propagation causes a nonlocal relation between local
temperature and heat insertion.  This paper focusses on the solution of the exact Peierls-Boltzmann equation (PBE), the relaxation time
approximation (RTA), and the definition of local temperature needed in both cases.  The concept of a non-local
``thermal susceptibility'' (analogous to charge susceptibility) is defined.  A formal solution is obtained for heating
with a single Fourier component $P(\vec{r},t)=P_0 \exp(i\vec{k}\cdot\vec{r}-i\omega t)$, where $P$ is the local rate of heating).
The results are illustrated by Debye model calculations in RTA for a three-dimensional periodic system where heat is added 
and removed with $P(\vec{r},t)=P(x)$ from isolated evenly spaced segments with period $L$ in $x$.  
The ratio $L/\ell_{\rm min}$ is varied from 6 to $\infty$, where $\ell_{\rm min}$ is the minimum mean free path.
The Debye phonons are assumed to scatter anharmonically with mean free paths varying as $\ell_{\rm min}(q_D/q)^2$
where $q_D$ is the Debye wavevector.  The results illustrate the expected local (diffusive) response for $\ell_{\rm min}\ll L$,
and a diffusive to ballistic crossover as $\ell_{\rm min}$ increases toward the scale $L$.  The results also illustrate the
confusing problem of temperature definition.  This confusion is not present in the exact treatment but occurs in RTA.
\end{abstract}

\maketitle


\section{Introduction}

Phonons have diverse mean free paths $\ell_Q$, diverging as $\omega_Q^{-p}$ at low frequencies.  The power $p$ is 
also diverse.  For anharmonic scattering, computations \cite{Esfarjani2011,Ma2014,Zhou2016} 
and experiment \cite{Damen1999}
confirm Herring's \cite{Herring1954} predictions: for scattering by N (normal) processes, $p=2$ and 
by U (Umklapp) processes, $p=3$.
For scattering by defects (Rayleigh scattering) $p=4$.  This diversity is revealed as nonlocality
(often designated as the ``ballistic/diffusive crossover'') in the relation between phonon heat current and temperature
or temperature gradient.  Early discussions of nonlocality ({\it e.g.} ref. \onlinecite{Simons1960})
considered transport in media bounded only perpendicular to the direction of the current.
The more fundamental non-locality 
for finite size parallel to the current has been discussed for phonon transport since
at least 1980 \cite{Levinson1980,Mahan1988,Majumdar1993,Chen2005,Shinde2014}.
It causes interesting complexities at submicron length scales \cite{Chen2005,Fisher2014,Carlos2016}, 
currently a topic under intense study.
This paper uses phonon quasiparticle theory as described by the Peierls-Boltzmann equation (PBE) \cite{Peierls1929}.
The PBE is actually $3nN$ equations, one for the distribution $N_Q$ of each phonon mode $Q=(\vec{q},s)$, where $\vec{q}$ is
one of the $N$ wavevectors of the crystal with $N$ unit cells, and $s$ runs over the $3n$ branches.

The quasiparticle distribution $N_Q$ is driven by external heating at a rate $P(\vec{r},t)$.
This driving is described by a term in the PBE which has only recently been discussed 
\cite{Hua2014b,Ramu2014,VermeerschI2015,Hua2018}.  The heating $P$
causes the temperature to have new spatial variation $T=T_0 +\Delta T(\vec{r},t)$.  Thus there are 
$3nN+2$ fields, $N_Q(\vec{r},t)$, $P(\vec{r},t)$, and $\Delta T(\vec{r},t)$.  Typically $P$ is predetermined,
and $\Delta T$ should be calculated from $N_Q$ and $P$.  A new equation must be added to the
$3nN$ PBE's, in order to solve for $3nN+1$ unknown functions in terms of the given function $P$.  

For the conventional bulk problem, everything is homogeneous in space and time.  The heating
$P$ is distant from the region of interest, but has created a known heat current $\vec{j}$.
The temperature is $T_0$ plus a constant gradient which can be measured.  The thermal 
conductivity (the ratio of current to temperature gradient) can be computed from the PBE 
without need for an extra equation, since $P$ is irrelevant, $\vec{\nabla}T$ is given, and $\vec{j}$ is found from $N_Q$.
It is now common to implement a ``first-principles'' anharmonic phonon theory \cite{Tadano2014,Togo2015}.
Codes that permit full inversion of the PBE \cite{Li2014,Chernatynskiy2015} are widely accessible.  
The results for simple semiconductors \cite{Lindsay2013} are very impressive.
Similar computations for spatially inhomogeneous situations are starting to emerge \cite{Cepellotti2017,Carrete2017}.
PBE treatments using RTA are challenging enough.  A good example is Ref. \onlinecite{Hua2018}.

In the inhomogeneous case, measurement of
$\Delta T(\vec{r},t)$ is a challenge only partly solved, making computation (also only partly solved)
an important issue.  One object of this paper is to clarify what the additional equation should be.    For the
correct Boltzmann equation with energy conserved in microscopic phonon collisions, there is a
single sensible answer, namely that the local energy density $U(\vec{r},t)$ contained in the
distribution functions $N_Q$ should also be completely described by the local equilibrium distribution $n_Q$,
a Bose-Einstein distribution with the local temperature $T(\vec{r},t)$.  We regard this as a definition
of temperature in a spatially inhomogeneous situation.  This definition seems compulsory within the
context of kinetic theory.  There should be no net
local heat in the deviation $\Phi_Q=N_Q-n_Q$, even though $\Phi_Q$ contains all the 
heat current.  Unfortunately, this ``exact'' energy conserving PBE is very challenging to solve in a 
nonlocal situation, and the relaxation time approximation (RTA) is a desirable shortcut.  There
are two plausible candidates for the additional equation needed,
when RTA is used.  Neither is perfect.  They cannot both be satisfied.  They provide two alternative
definitions of temperature, which can be regarded as a shortcoming of RTA when applied to
spatially inhomogeneous situations.   The first alternative is to use the same condition needed in
the ``exact'' (meaning no RTA is used) theory.  Surprisingly, this choice seems more
problematic than the second definition, which arises by forcing the solution of the RTA-PBE to have
no net energy changes caused by collisions.  Energy conservation is strictly obeyed by each collision in the exact theory, 
and disobeyed in each collision in RTA.  However, it is sensible to force it to be true on average in RTA.   
We compare the results of these two candidate extra conditions
for a Debye model treated in RTA, and containing diverse phonon mean free paths.

It is convenient to formulate the theory in Fourier space, where the distribution function is $N_Q(\vec{k},\omega)$.
We will not use different notations for functions like $N_Q$ when they are in coordinate or reciprocal space.
The symbol $N_Q$ will mean the distribution function, which can be in either real or reciprocal space representation.
Specific variables $(\vec{r},t)$ or $(\vec{k},\omega)$ will often be omitted unless a particular
representation is being considered.  For our numerical work using a Debye and RTA model,
we will drop the time dependence.  Variables will depend of $\vec{r}$ or $\vec{k}$, but not on
$t$ or $\omega$.  But in the formal theoretical treatment, there is virtue in keeping
time dependence as an option.

\begin{figure}
\includegraphics[angle=0,width=0.4\textwidth]{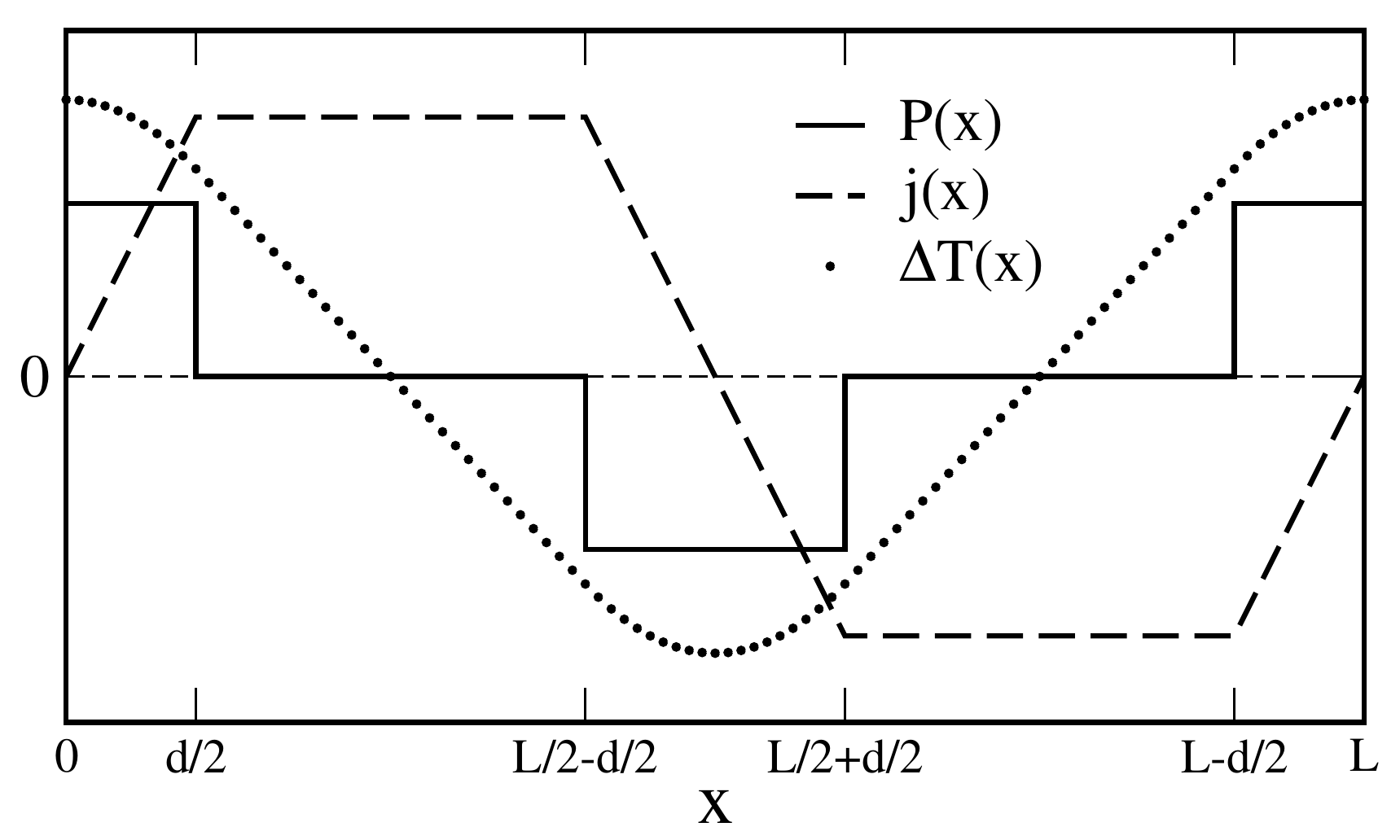}
\caption{\label{fig:figgrid} One period $0\le x \le L$ of a model periodic system
(to be studied numerically in Sec. \ref{sec:DMC}).  Static heating $P(x)$ is done in segments of length $d$
centered at $x=nL$.  Equal removal of heat is done in similar segments located at $x=(n+1/2)L$.  Heat currents $j$
obeying $\vec{\nabla}\cdot \vec{j}=P$ carry heat from hotter to colder regions.  The temperature excursion
$T(x)-T_0=\Delta T(x)$ is shown in the local (or diffusive) limit, where $\kappa d^2 T/dx^2 =-P(x)$.
Non-local effects on $T(x)$ are
expected if $x$ is within phonon mean free paths of regions of spatial variation of $P$.
Heating $P$, current $j$, and temperature $\Delta T$ are all in arbitrary units.
}
\end{figure}

Suppose heat is supplied to an insulating solid at rate $P(\vec{r},t)$, 
where $\int d\vec{r}P(\vec{r},t)=0$.   This guarantees that after transients have died, a steady state exists
with heat removal exactly compensating heat addition.
A model (and time-independent) example is shown in Fig. \ref{fig:figgrid}.
We assume that $P$ is a small perturbation, which allows a linear approximation.
The PBE, which governs the evolution of $N_Q$, becomes a linear equation, to be solved for
$\Delta T$ and $\Phi_Q$ to linear order in the driving $P$.

\par
\par

We should acknowledge that there is no uniquely accepted definition of temperature for systems out of equilibrium.  Rigorous
thermodynamics may even reject the attempt \cite{Elser2017}.  Attempts at general theories
are available \cite{Hill2001,Chamberlin2015,Stafford2016}.  We have two remarks. (1)
If a measurement can be well interpreted in terms of a $\Delta T$, this represents for us a sufficient
definition for that problem.  (2) A careful Boltzmann treatment with a correct quasiparticle
scattering operator necessarily introduces a local temperature $T(\vec{r},t)$; this is the object of our study.
Many schemes for measuring $\Delta T(\vec{r},t)$ have been devised and used \cite{Brites2012,Cahill2014,Reparaz2014};
for example, transient thermal gratings \cite{Vega-Flick2016} or stationary physical gratings \cite{Zeng2015}.
Molecular dynamics (MD) modelling \cite{Zhou2009,McGaughey2014,Wang2017}
also provides ``data'' of this type, and Monte Carlo simulations \cite{Peraud2011,Peraud2014} are useful.

\section{Boltzmann equation}

If phonons are the only heat carrier, space and time variations not too rapid, and scattering not so strong
as to degrade the phonon quasiparticle picture, then the PBE applies,
\begin{equation}
\frac{\partial N_Q}{\partial t}=\left(\frac{\partial N_Q}{\partial t}\right)_{ \rm drift}
+ \left(\frac{\partial N_Q}{\partial t}\right)_{ \rm scatt}
+\left(\frac{\partial N_Q}{\partial t}\right)_{\rm ext}
\label{eq:PBEQ}
\end{equation}
It is convenient to have a vector-space notation, where $|Q\rangle$ is a $3nN$-vector containing
the components of the normal mode eigenvectors and $|N\rangle$ is the distribution function.
Its normal mode component is $N_Q=\langle Q|N\rangle$.  In this notation, Eq. \ref{eq:PBEQ} is
\begin{equation}
\frac{\partial}{\partial t}\left| N\right\rangle=\frac{\partial}{\partial t}\left|N\right\rangle_{ \rm drift}
+ \frac{\partial}{\partial t}\left|N\right\rangle_{ \rm scatt}
+\frac{\partial}{\partial t}\left|N\right\rangle_{\rm ext}
\label{eq:PBEv}
\end{equation}
This notation appears occasionally in the literature, {\it e.g.} Ref. \onlinecite{Guyer1966}.
The phonon energy density $U(\vec{r},t)$ or $U(\vec{k},\omega)$ is
\begin{equation}
U=\frac{E}{V}=\frac{1}{V}\sum_Q \hbar\omega_Q N_Q = \frac{1}{V} \langle \hbar\Omega|N\rangle,
\label{eq:U}
\end{equation}
where $V$ is the volume of the crystal.
To clarify the compact vector notation, note that the unit operator can be written in normal
mode space as $1=\sum_Q |Q\rangle\langle Q|$.  Then $U=(1/V)\sum_Q\langle \hbar\Omega|Q\rangle
\langle Q|N\rangle$.
The inner product $\langle \hbar\Omega|Q\rangle$ is just $\hbar\omega_Q$.  

In the full Boltzmann equation \cite{Ziman1960}, 
the scattering term $(\partial N_Q/\partial t)_{ \rm scatt}$ is a complicated non-linear
function of the distributions $N_{Q^\prime}$.  It conserves phonon energy but, because of Umklapp
processes, crystal momentum $\vec{q}$ is not conserved.  Boltzmann's H-theorem tells us \cite{Pauli1928} that
collisions cannot decrease entropy, only increase it, where nonequilibrium entropy is defined for phonons as
$S/k_B = \sum_Q [ (N_Q+1)\ln(N_Q+1)-N_Q\ln N_Q ]$.  Entropy stops increasing when it reaches the maximum
consistent with the available local phonon energy.  This maximum occurs when $N_Q$ evolves to the Bose
distribution $n_Q(T(\vec{r},t))$, where the definition of local temperature $T(\vec{r},t)$ is that 
\begin{equation}
U = \frac{1}{V} \langle \hbar\Omega|N\rangle =  \frac{1}{V} \langle \hbar\Omega|n(T(\vec{r},t))\rangle.
\label{eq:UU}
\end{equation}
%
The distribution function can be written as $N_Q=n_Q(T(\vec{r},t)) + \Phi_Q$, or equivalently 
$|N\rangle=|n(T(\vec{r},t)\rangle+|\Phi\rangle$, where the local temperature is the one that satisfies Eq. \ref{eq:UU}.
Then the scattering term in the Boltzmann equation, after linearizing in $\Phi_Q$, takes the form 
\begin{equation}
\left(\frac{\partial N_Q}{\partial t}\right)_{ \rm scatt}=-\sum_{Q^\prime} S_{QQ^\prime}\Phi_{Q^\prime} \ \ {\rm or} \ \ 
\frac{\partial}{\partial t}|N\rangle_{ \rm scatt}=-S|\Phi\rangle,
\label{eq:scfull}
\end{equation}
where $S_{QQ^\prime}=\langle Q|S|Q^\prime\rangle$.  The deviation 
 $\Phi_Q$ transports heat but can have no net energy, $\sum_Q\hbar\omega_Q\Phi_Q=\langle \hbar\Omega|\Phi\rangle=0$.
There is nothing in the Boltzmann equation itself that can specify the value of $T(\vec{r},t)$.  The correct
specification is just the extra constraint $\langle \hbar\Omega|\Phi\rangle=0$ that has to be imposed.
 
When time-independent bulk thermal conductivity is
studied, one ignores the details of heat addition and removal at distant places, and instead
assumes that $T(\vec{r})$ equals the background temperature $T_0$ plus a small correction $\Delta T$ with a 
constant gradient $\vec{\nabla}T$.  Then one
solves the PBE to find the resulting constant $\vec{j}$.  However, we need to deal with cases where
the known quantity is the heat input, and $T(\vec{r},t)$ is unknown.

Phonon energy is conserved in collisions,
\begin{equation}
\left(\frac{\partial U}{\partial t}\right)_{ \rm scatt}=0=-\frac{1}{V}\langle \hbar\Omega |S| \Phi\rangle.
\label{eq:EconS}
\end{equation}
This equation is satisfied for any deviation $|\Phi\rangle$.  This is equivalent to the statement that the dual-space vector 
$\langle \hbar\Omega|$ is a null left eigenvector of the linearized scattering operator,
$\langle \hbar\Omega|S=0$.  This can be shown explicitly using standard \cite{Ziman1960} third-order anharmonic scattering.

Linear approximation allows separate treatment of each Fourier component.
Defining, for example, 
\begin{equation}
\Phi_Q(\vec{r},t)=\frac{1}{2\pi}\int_{-\infty}^\infty d\omega \sum_{\vec{k}} \Phi(\vec{k},\omega)e^{i(\vec{k}\cdot\vec{r}-\omega t)}
\label{eq:FT}
\end{equation}
the drift term $(\partial N_Q/\partial t)_{ \rm drift}=-\vec{v}_Q \cdot\vec{\nabla} N_Q$ has the form
\begin{equation}
\left(\frac{\partial N_Q (\vec{k},\omega)}{\partial t}\right)_{ \rm drift}=-i\vec{k}\cdot\vec{v}_Q \frac{dn_Q}{dT}\Delta T(\vec{k},\omega)
-i\vec{k}\cdot\vec{v}_Q  \Phi_Q (\vec{k},\omega),
\label{eq:dr}
\end{equation}
where $\vec{v}_Q=\partial\omega_Q/\partial\vec{q}$ is the phonon group velocity.
In vector notation this is
\begin{equation}
\frac{\partial}{\partial t} \left| N(\vec{k},\omega)\right\rangle_{ \rm drift} 
= -i\vec{k}\cdot\vec{v} \left| \frac{dn}{dT} \right\rangle\Delta T(\vec{k},\omega)
-i\vec{k}\cdot\vec{v}|\Phi(\vec{k},\omega)\rangle
\label{eq:fdr}
\end{equation}
where each component $v_\alpha$ is a $3nN\times 3nN$ matrix, diagonal in the normal mode representation,
\begin{equation}
\langle Q|\vec{v}|Q^\prime\rangle=\vec{v}_Q \delta_{Q,Q^\prime}.
\label{eq:v}
\end{equation}

Finally, the PBE needs a term $(\partial N_Q/\partial t)_{\rm ext}$ which describes how external heat
(at rate $P(\vec{r},t)$ or $P(\vec{k},\omega)$)
is added and removed to keep a steady-state inhomogeneous temperature and
heat current.  It seems unlikely that there is a single
universal form.  Hua and Minnich \cite{Hua2014b} use a somewhat more general
form than the one used here, which was introduced by Vermeersch {\it et al.} \cite{VermeerschI2015},
\begin{equation}
\left(\frac{\partial N_Q }{\partial t}\right)_{\rm ext}=\frac{P}{C} \frac{dn_Q}{dT}.
\label{eq:extQ}
\end{equation}
The idea is that added heat causes a time rate of increase of occupancy $N_Q$ of mode $Q$,
identical to what would happen close to equilibrium with a
time rate of temperature increase $P/C$.  Here $C$ is the bulk specific heat.
This equation does not correspond closely to any particular experiment.  It does
agree with typical molecular dynamics (MD) simulations which use local thermostatting,
for example, ref. \onlinecite{Zhou2009}.  In vector notation,
\begin{equation}
\frac{\partial}{\partial t}\left| N\right\rangle_{\rm ext}=\frac{P}{C}\left|\frac{d n}{dT}\right\rangle.
\label{eq:ext}
\end{equation}
The specific heat $C=\langle  \hbar\Omega|dn/dT\rangle/V=\sum_Q C_Q$
is the sum of contributions $C_Q$ from each normal mode,
\begin{equation}
C_Q=\frac{1}{V}\hbar\omega_Q\frac{dn_Q}{dT}.
\label{eq:CQ}
\end{equation}

Now we can write the ``full'' (meaning not RTA) linearized PBE,
\begin{equation}
\frac{\partial}{\partial t}| N(\vec{r},t)\rangle =\left(\frac{P}{C} -\vec{v}\cdot\vec{\nabla}T\right) \left|\frac{dn}{dT}\right\rangle
-(S+\vec{v}\cdot\vec{\nabla}) |\Phi\rangle 
\label{eq:fullPBErt}
\end{equation}
\begin{equation}
(S+i\vec{k}\cdot\vec{v}-i\omega) |\Phi\rangle
=\left(\frac{P}{C} -i ( \vec{k}\cdot\vec{v}-\omega )\Delta T\right) \left|\frac{dn}{dT}\right\rangle
\label{eq:fullPBEkw}
\end{equation}
All fields $N, \ n, \ \Phi, \ P, \ {\rm and} \ \Delta T$ are in real space $(\vec{r},t)$ in Eq. 
\ref{eq:fullPBErt}, or Fourier space $(\vec{k},\omega)$ in Eq. \ref{eq:fullPBEkw}.  
The Fourier space version is simpler because it gets rid
of  differential operators $\partial/\partial t$ and $\vec{\nabla}$.  Taking the projection onto mode $Q$,
{\it i.e.} operating on the left by $\langle Q|$, we have $3nN$ equations, one for each $Q$, that can
be solved for $\Phi_Q$ in terms of the fields $P$ and $\Delta T$. We wish to apply this to 
problems where $P$ is given.  Then $\Delta T$ needs to be specified by an additional equation,
already suggested by the H-theorem, and discussed in the next section.

\section{Energy and energy conservation}

\subsection{Full treatment}

The total quasiparticle energy density $U(\vec{r},t)$ is defined in Eq. \ref{eq:U}.
The time rate of change is $\partial U/\partial t=
\langle  \hbar\Omega|\partial N/\partial t\rangle/V$.  Taking the $(1/V)\langle \hbar\Omega|$
projection of Eq. \ref{eq:fullPBErt}, the
left side gives $\partial U/\partial t$. The first part of the first term on the right is just $P_0$, because 
$\langle  \hbar\Omega | dn/dT\rangle/V$ is the specific heat $C$. 
The second part of the first term on the right vanishes because time-reversal
symmetry requires $\langle  \hbar\Omega|\vec{v}|n\rangle=0$.  The first part of the second term on the right vanishes
because $\langle \hbar\Omega|S|\Phi\rangle/V=(\partial U/\partial t)_{\rm coll}=0$ is the statement that
collisions do not change the total energy.  The second part of the second term on the right is $-\vec{\nabla}\cdot\vec{j}$,
where
\begin{equation}
\vec{j}=\frac{1}{V}\langle \hbar\Omega|\vec{v}|\Phi\rangle=\frac{1}{V}\sum_Q \hbar\omega_Q \vec{v}_Q \Phi_Q.
\label{eq:j}
\end{equation}
Putting it together, the answer is
\begin{equation}
\frac{\partial U(\vec{r},t)}{\partial t}=-\vec{\nabla}\cdot\vec{j}(\vec{r},t)+P(\vec{r},t),
\label{eq:Econsv}
\end{equation}
or, the rate of local energy increase is the sum of energy current flowing in and external heating.

\subsection{Relaxation time approximation}

Consider first how a phonon quasiparticle relaxes toward equilibrium.  Suppose that mode $Q$ is
the only mode not in equilibrium, which means $\Phi_{Q^\prime}=\Phi_Q \delta_{Q,Q^\prime}$.
Then Eq. \ref{eq:scfull} reduces to
\begin{equation}
\left(\frac{\partial N_Q}{\partial t}\right)_{\rm relax} = -\frac{N_Q - n_Q}{\tau_Q}
\label{eq:RT}
\end{equation}
where $1/\tau_Q = S_{QQ}$ is the quasiparticle relaxation rate.  The rate $1/\tau_Q$ is therefore
the mode-diagonal element of the operator $S$.  It is often
called the ``single mode relaxation rate''.  Solving the full PBE, Eq. \ref{eq:fullPBEkw} is challenging because
one needs to invert a large non-Hermitean matrix $S+i\vec{k}\cdot\vec{v}-i\omega$ for many $\vec{k}$'s. 
The only implementation we know of is Ref. \onlinecite{Cepellotti2017}.  The problem is greatly simplified if
off-diagonal elements $S_{QQ^\prime}$ of the mode-space scattering matrix are ignored.  This
is the RTA, $S\rightarrow S_D$, where $\langle Q|S|Q^\prime\rangle$ is approximated by its $Q$-diagonal matrix 
elements $S_{D,QQ^\prime}=1/\tau_Q \delta_{QQ^\prime}$.  The notation $S_D$ denotes
the diagonal part of $S$.  This means using Eq. \ref{eq:RT}
for the scattering term in the PBE.  Unfortunately, this destroys energy conservation.
It is known that, when $\vec{k}=0$, this approximation is often quite good, but little is known about the 
accuracy of RTA in the inhomogeneous case.  
For bulk thermal conductivity ($\vec{k}=0$), $\vec{k}$ and $\vec{\nabla}\Phi$ go to zero, and the matrix
to be inverted is $S$ rather than $S+i\vec{k}\cdot\vec{v}$.  This greatly simplifies the problem. 
``First-principles'' calculations doing full inversion of $S$ compare very well with the RTA use of only 
diagonal parts of $S$, for simple semiconductors at $T$ not too low \cite{Lindsay2013}.  The accuracy
of RTA for $\vec{k}\ne 0$ calculations has not been similarly tested.

\section{Solution of the nonlocal PBE}

The formal solution of Eq. \ref{eq:fullPBEkw} is
\begin{equation}
|\Phi\rangle=(S+i\vec{k}\cdot\vec{v}-i\omega)^{-1}\left(\frac{P}{C}
-(i\vec{k}\cdot\vec{v}-i\omega)\Delta T \right)\left|\frac{dn}{dT}\right\rangle.
\label{eq:Fsoln}
\end{equation}
The deviation $\Phi$ has a piece driven by $P$ and another driven by $\Delta T$.  As
argued above, an extra equation is needed, namely $\langle  \hbar\Omega| \Phi\rangle =0$.
This is the simplest sensible definition of a non-equilibrium temperature.

\subsection{Nonlocal Thermal susceptibility}

Using $\langle  \hbar\Omega| \Phi\rangle =0$, the relation between $P$ and $\Delta T$ is
\begin{equation}
C(T)\Delta T(\vec{k},\omega) = \Theta(\vec{k},\omega) P(\vec{k},\omega)
\label{eq:defTheta}
\end{equation}
where
\begin{equation}
\Theta(\vec{k},\omega) = \frac
{\left\langle \hbar\Omega\left|(S+i\vec{k}\cdot\vec{v}-i\omega)^{-1} \right|\frac{dn}{dT}\right\rangle}
{\left\langle \hbar\Omega\left|(S+i\vec{k}\cdot\vec{v}-i\omega)^{-1} 
(i\vec{k}\cdot\vec{v}-i\omega)\right|\frac{dn}{dT}\right\rangle}
\label{eq:Theta}
\end{equation}
Equation \ref{eq:defTheta} defines the nonlocal ``thermal susceptibility'' $\Theta$ as the temperature
response to external heat input.  It is interesting to consider this first, before going to the nonlocal thermal conductivity
$\kappa(\vec{k},\omega)$.  Generalization to
spatial and temporal inhomogeneity is considered natural for the electrical conductivity $\sigma(\vec{k},\omega)$.
Unlike the electrical case where driving
is caused by a well-defined $E$-field, the $P$-field causing inhomogeneous thermal response
does not have a unique form.  In addition, the notions of local heat current \cite{Hardy1963,Ercole2016} and
local temperature gradient are both somewhat insecure.  
The scalar $\Theta$ is perhaps more relevant  than the tensor $\kappa$ to non-local heat evolution.  The temperature
$\Delta T(\vec{r},t)$ is difficult to measure; appropriate theoretical assistances would help.  
Both $\Theta$ and $\kappa$ are causal;
$\Delta T(t)$ does not respond to $P(t^\prime)$ unless $t^\prime <t$, nor does $\vec{j}(t)$ respond
to $\vec{\nabla}T(t^\prime)$ unless $t^\prime <t$.  Local energy density and temperature are
questionable concepts if sources $P(t)$ vary rapidly in time.  $\Theta(\omega)$ and
$\kappa(\omega)$ are probably useful only at low $\omega$, smaller than typical $\omega_Q$'s,
but not smaller than $1/\tau_Q$'s for the longer-lived phonons.

\subsection{Thermal Conductivity}

Now compute the heat current using Eq. \ref{eq:j}.
Eliminating $P_0/C$ in favor of $\Delta T$, by use of Eqns. \ref{eq:defTheta} and \ref{eq:Theta}, gives
\begin{equation}
\vec{j}=\frac{1}{V}\langle \hbar\Omega|\vec{v} \ \Xi^{-1}\left(\frac{1}{\Theta}-i\vec{k}\cdot\vec{v}+i\omega\right)
\left|\frac{dn}{dT}\right\rangle \Delta T
\label{eq:j1}
\end{equation}
where the shorthand is introduced,
\begin{equation}
\Xi=S+i\vec{k}\cdot\vec{v}-i\omega.
\label{eq:Xi}
\end{equation}
Notice that from Eq.(\ref{eq:Theta}) $1/\Theta$ can be written as
\begin{equation}
\frac{1}{\Theta}= -i\omega + \frac{\langle \hbar\Omega|\Xi^{-1}(i\vec{k}\cdot\vec{v})\left|\frac{dn}{dT}\right\rangle}
{\langle \hbar\Omega|\Xi^{-1}\left|\frac{dn}{dT}\right\rangle}.
\label{eq:theta1}
\end{equation}
The $i\omega$ terms now cancel from $(\cdot )$ in Eq. \ref{eq:j1}, leaving the expression
$i\vec{k}\cdot\vec{v}$ in both remaining parts of $(\cdot)$.  The factor $i\vec{k}$ can be taken
outside the $\langle\cdot\rangle$ elements and combined with $\Delta T$, which is then
rewritten as  $i\vec{k}\Delta T = \vec{\nabla}T$.
Then the current (Eq. \ref{eq:j}) is $\vec{j}=-\mathbf{\kappa}\cdot\vec{\nabla}T$,
where the conductivity tensor is
\begin{eqnarray}
\mathbf{\kappa}(\vec{k},\omega)&=&\frac{1}{V}\left\langle \hbar\Omega\left|\vec{v} \ \Xi^{-1}
\vec{v}\right| \frac{dn}{dT}\right\rangle \nonumber \\
&-& \frac{1}{V}\frac{\left\langle \hbar\Omega\left|\vec{v} \ \Xi^{-1}
\right|\frac{dn}{dT}\right\rangle
\left\langle \hbar\Omega\left| \Xi^{-1}\vec{v}\right|\frac{dn}{dT}\right\rangle}
{\left\langle \hbar\Omega\left| \Xi^{-1}\right|\frac{dn}{dT}\right\rangle} \nonumber \\
\label{eq:Fkappa}
\end{eqnarray}
In the static ($\omega\rightarrow 0$) homogeneous ($\vec{k}\rightarrow 0$) limit, 
$\Xi$ becomes $S$.  The second term of Eq. \ref{eq:Fkappa} then
vanishes by time-reversal symmetry, and the answer becomes
\begin{equation}
\kappa_{\alpha\beta}(\vec{k}=0,\omega=0)=\frac{1}{V}\left\langle \hbar\Omega\left|v_\alpha S^{-1}v_\beta \right|\frac{dn}{dT}\right\rangle.
\label{eq:Fkappa0}
\end{equation}
This is the solution of the standard PBE ({\it i.e.} $\kappa\approx Cv\ell/3$) for bulk thermal conductivity.

\subsection{Extracting $\Delta T$ from $P$}

There are two ways to find the unknown inhomogeneous temperature $\Delta T(\vec{k},\omega)$,
which can then be Fourier transformed to $\vec{r},t$.  The direct route is from the
thermal susceptibility, Eq. \ref{eq:defTheta}.  The less direct route is to use the known current $\vec{j}$ and the
non-local conductivity $\kappa$ (Eq. \ref{eq:Fkappa}) to find the temperature gradient $i\vec{k}\Delta T$.
In an approximate theory (like the RTA) these routes do not necessarily give identical results.  Here are
three versions of temperature:
\begin{equation}
P(\vec{k},\omega)=C\Delta T_\theta(\vec{k},\omega) /\Theta(\vec{k},\omega)
\label{eq:T1}
\end{equation}
\begin{equation}
i\vec{k}\cdot\vec{j}(\vec{k},\omega)=-(i\vec{k})\cdot \kappa(\vec{k},\omega)\cdot(i\vec{k}) \Delta T_\kappa (\vec{k},\omega)
\label{eq:T2}
\end{equation}
\begin{equation}
i\vec{k}\cdot\vec{j}(\vec{k},\omega)=P(\vec{k},\omega) +i\omega C \Delta T_U (\vec{k},\omega)
\label{eq:T3}
\end{equation}
These three versions are labeled $\Delta T_\theta$, $\Delta T_\kappa$, and $\Delta T_U$ because they derive from
thermal susceptibility $\Theta$ (Eq. \ref{eq:defTheta}), thermal conductivity $\kappa$ (Eq. \ref{eq:Fkappa}), and
energy conservation (Eq. \ref{eq:Econsv}).  If all three are equal, we can combine the equations to get 
\begin{equation}
-(i\vec{k})\cdot \kappa(\vec{k},\omega)\cdot(i\vec{k})=\left(\frac{1}{\Theta(\vec{k},\omega)}+i\omega\right)C
\label{eq:xSx}
\end{equation}
This equation is indeed satisfied by the exact formal solutions Eq. \ref{eq:Fkappa} and \ref{eq:Theta}.
It is reassuring to know that all three versions of $T(\vec{r},t)$ are the same according to the PBE.
Equation \ref{eq:xSx} is about the longitudinal part of the thermal
conductivity, $\kappa_L = (\Theta(\vec{k},\omega)^{-1} +i\omega)C/k^2$.  This is reminiscent of the formalism for electrical response \cite{Dressel2002}.  
The conductivity tensor $\sigma$ is a causal current-current response function, and relates directly to the
dielectric tensor $\epsilon=1+4\pi i \sigma/\omega$.  The longitudinal dielectric response has a reciprocal
relation to a susceptibility, similar to Eq. \ref{eq:xSx}, namely $\epsilon_L^{-1}=1+v(k)\chi(\vec{k},\omega)$, where the susceptibility
$\chi$ is the causal charge density-charge density response function \cite{Pines1966,Giuliani2005}.  The thermal susceptibility
$\Theta$ seems analogous to the electrical susceptibility $\chi$.

\section{RTA version of Full Solution}

The RTA is the approximation of keeping only the diagonal terms $\langle Q|S_D|Q\rangle=1/\tau_Q$
of the full linearized scattering operator $S_{Q,Q^\prime}=\langle Q|S|Q^\prime\rangle$.
RTA formulas can be derived in two equivalent ways.  (1) Take the full solution Eq. \ref{eq:Fkappa} and replace $S$ by $S_D$.
(2) Use the RTA version of the PBE, Eq. \ref{eq:fullPBEkw} with $S\rightarrow S_D$, and supplement it with the
RTA version of $\langle  \hbar\Omega|\Phi\rangle=0$.
Both methods generate the same anwers for $\Theta$ and $\kappa$. However, because scattering
in RTA does not conserve energy
in collisions, Eq. \ref{eq:xSx} is not obeyed by the resulting approximate $\Theta$ and $\kappa$.
The RTA version of Eq. \ref{eq:Fkappa} is labeled ``$\kappa_{\rm RTA,A}$'' because a ``B'' version will soon be discussed.
\begin{eqnarray}
&&\mathbf{\kappa}_{\rm RTA,A}^{\alpha\beta}(\vec{k})=\sum_Q \frac{ C_Q
v_{Q\alpha}v_{Q\beta}}{1/\tau_Q+i\vec{k}\cdot\vec{v}_Q }
 \nonumber \\ &+&
\frac{ \sum_Q \frac{ C_Q (iv_{Q\alpha}) } {1/\tau_Q+i\vec{k}\cdot\vec{v}_Q }
\sum_{Q^\prime} \frac{ C_{Q^\prime}
(iv_{Q^\prime \beta}) } {1/\tau_{Q^\prime}+i\vec{k}\cdot\vec{v}_{Q^\prime} }}
{\sum_{Q^{\prime\prime}} \frac{ C_{Q^{\prime\prime}} } 
{1/\tau_{Q^{\prime\prime}}+i\vec{k}\cdot\vec{v}_{Q^{\prime\prime}}  } }
\label{eq:RTAkappa}
\end{eqnarray}
This has been written for the case $\omega=0$.  For non-zero $\omega$, simply
replace $\vec{k}\cdot\vec{v}$ by $\vec{k}\cdot\vec{v}-\omega$.
The integrals in both numerator and denominator of the second term are real and positive,
so the second term (for diagonal elements $\kappa_{\alpha\alpha}$ is a positive correction to the first.
In the local limit $\vec{k}\rightarrow 0$, the second term disappears and the answer has the familiar
form $\kappa_{\alpha\beta}=\sum_Q C_Q v_{Q\alpha} v_{Q\beta} \tau_Q$  The second term,
the correction coming from spatially inhomogeneous driving, gives the dominant contribution
when $|\vec{k}|\ell_Q\ge 1$.

The first term of Eq. \ref{eq:RTAkappa} is not at all surprising.  It is a close analog of the
usual formula for the nonlocal Drude conductivity $\sigma(\vec{k},\omega)$ of a metal.
This is the formula of Reuter and Sondheimer \cite{Reuter1948,Grosso2000},
which clarified Pippard's theory \cite{Pippard1947} of the anomalous skin effect.
The analogous electronic RTA formula is
\begin{equation}
\sigma_{xx}(\vec{k},\omega) = \frac{e^2}{V}\sum_Q \frac{v_{Qx}^2}{1/\tau_Q+i\vec{k}\cdot\vec{v}_Q
-i\omega} \frac{\partial f_Q}{\partial \mu}.
\label{eq:elec}
\end{equation}
Here the index Q labels the electron Bloch states of energy $\epsilon_Q$, group velocity $\vec{v}_Q$,
and equilibrium Fermi-Dirac occupancy $f_Q$, and $\mu$ is the chemical potential.

\section{Alternate solution starting from RTA}

Equation \ref{eq:RTAkappa} does not agree with previous RTA solutions found in the literature 
\cite{Vermeersch2014,Hua2015}.   The reason is that there is an
alternative constraint that competes with $\langle \hbar\Omega|\Phi\rangle=0$, namely, instead of choosing
$\Delta T$ so that $\Phi_Q$ contains no net energy, $\Delta T$ can be chosen so that
the RTA collisions conserve energy on average.  These are both valid and desirable constraints, but
they are not compatible and cannot both be satisfied in RTA.  The two competing constraints are:

{\bf A}: Route {\bf A} is the same as the one required in the full PBE: $\sum_Q \hbar\omega_Q \Phi_Q=0=\langle \hbar\Omega|\Phi\rangle$.

{\bf B}: Route {\bf B} forces the RTA scattering term to be energy conserving: $\sum_Q \hbar\omega_Q\Phi_Q/\tau_Q=0
=\langle \hbar\Omega|S_D|\Phi\rangle$.

The solution without RTA {\it via} route {\bf A} gives the thermal susceptibility $\Theta=C\Delta T/P$ of Eq. \ref{eq:Theta}.
When RTA is used, this becomes 
\begin{equation}
\Theta_{\rm RTA,A}=
\frac{\sum_Q \frac{C_Q}{C}\left(\frac{1}{1/\tau_Q + i\vec{k}\cdot\vec{v}_Q}\right)}
{\sum_Q\frac{C_Q}{C}\left(\frac{i\vec{k}\cdot\vec{v}_Q}{1/\tau_Q + i\vec{k}\cdot\vec{v}_Q}\right)}
\label{eq:ThetaA}
\end{equation}
Using this relation to compute $\vec{j}$ recovers Eq. \ref{eq:RTAkappa}.  The analog of route {\bf B},
$\langle \hbar\Omega|S|\Phi\rangle=0$, is not helpful for the full PBE (without RTA), because it is
automatically satisfied by the correct scattering operator $S$.    
However, route {\bf B} used in RTA is a sensible constraint, but gives a different formula for $\Theta$,
\begin{equation}
\Theta_{\rm RTA,B}=
\frac{\sum_Q \frac{C_Q}{C}\left(\frac{1/\tau_Q}{1/\tau_Q + i\vec{k}\cdot\vec{v}_Q}\right)}
{\sum_Q\frac{C_Q}{C}\left(\frac{i\vec{k}\cdot\vec{v}_Q /\tau_Q}{1/\tau_Q + i\vec{k}\cdot\vec{v}_Q}\right)}
\label{eq:ThetaB}
\end{equation}
For a given input power $P_0$, the two temperatures $T_A$ and $T_B$ are different.  When Eq. \ref{eq:ThetaB}
is used to compute $\vec{j}$, one gets the formulas for $\kappa(k)$ derived in Refs.
\onlinecite{Vermeersch2014} and \onlinecite{Hua2015},
\begin{eqnarray}
&&\mathbf{\kappa}_{\rm RTA,B}^{\alpha\beta}(\vec{k})=\sum_Q \frac{ C_Q
v_{Q\alpha}v_{Q\beta}}{1/\tau_Q+i\vec{k}\cdot\vec{v}_Q }
 \nonumber \\ &+&
\frac{ \sum_Q \frac{ C_Q (iv_{Q\alpha} )} {1/\tau_Q+i\vec{k}\cdot\vec{v}_Q }
\sum_{Q^\prime} \frac{ C_{Q^\prime}(i v_{Q^\prime \beta}) /\tau_{Q^\prime} } 
{1/\tau_{Q^\prime}+i\vec{k}\cdot\vec{v}_{Q^\prime} }}
{\sum_{Q^{\prime\prime}} \frac{ C_{Q^{\prime\prime}}/\tau_{Q^{\prime\prime}} } 
{1/\tau_{Q^{\prime\prime}}+i\vec{k}\cdot\vec{v}_{Q^{\prime\prime}}  } }.
 \nonumber \\
\label{eq:RTAkappaB}
\end{eqnarray}
Because of energy-conserving scattering, the route {\bf B} Eqs. \ref{eq:ThetaB} and \ref{eq:RTAkappaB}
do obey the relation Eq. \ref{eq:xSx}.
Notice that the first term in $\kappa_{\rm RTA}$ is identical for versions {\bf A} and {\bf B}.
In the ``gray model'' where $1/\tau_Q$ is (unrealistically) taken to be a constant $1/\tau$, all terms in
Eq. \ref{eq:RTAkappaB} and Eq. \ref{eq:RTAkappa} agree.
In the limit of small $k$ (the local or diffusive limit where Fourier's law applies), 
the first term in $\kappa_{\rm RTA, A \ or \ B}$ dominates and the two
versions agree.  In the opposite limit, the second term in $\kappa$ (coming from $\Theta$) dominates.  As will be seen
in the next section, results at large $k$ also show agreement between routes {\bf A} and {\bf B}.  
In the intermediate region, the two versions of $\kappa_{\rm RTA}$ disagree.
It would be interesting to compare first principles results using each of these equations, to see
which, if either, agrees well with the exact first principles result using Eq. \ref{eq:Fkappa}, but this is
beyond the scope of this paper.  

We should also mention that a referee has shown us
that route {\bf B} gives a more sensible answer to the diffusive 
thermal response of phonons to a point pulse perturbation.  
This adds weight to the argument in favor of route {\bf B} in RTA theory.

\section{Debye model calculations}
\label{sec:DMC}

To illustrate the differences between the versions {\bf A} and {\bf B},
we use the Debye model in three dimensions.  The three branches of phonons all have
$\omega_Q = v|\vec{q}|$ with the same velocity $v$, and scattering rate $1/\tau_Q=(1/\tau_D)(q/q_D)^2$.
The $\kappa_{\alpha\beta}$ tensors 
in Eqs. \ref{eq:RTAkappa} and \ref{eq:RTAkappaB} are scalars $\kappa\delta_{\alpha\beta}$.
The mean free path $\ell_Q=v\tau_Q$ has a minimum value $\ell_D=v\tau_D$.  The notations
[$\ell_{\rm min}$ and $\tau_{\rm min}$] are used interchangeably (in text and figures) with
[$\ell_D$ and $\tau_D$].  Debye model results are shown here in graphs.  Details of the formulas are 
discussed in the appendix.

There are three important length scales in the problem. (1) The shortest length scale $L_1$ 
is the lattice constant $a$, 
or the wavelength $\lambda_Q=2\pi/|\vec{ q}|$ of the short wavelength phonons. 
(2) Phonons have mean free paths $\ell_Q=\ell_D ( q_D/ q)^2$ in our Debye model .  
This gives a second length scale $L_2$, namely
$L_2= \ell_{\rm min}(T)$, the temperature-dependent minimum mean free path.
(3) The length scale $L_3$ characterizes the spatial variation of $|\vec{\nabla}T|$.
This scale is determined by sample
and heater geometry, {\it i.e.} how close to the heater are we interested to know the spatial variation of temperature
$T(\vec{r})$.  This spatial variation determines the shorter 
important wavevectors $k^\ast \approx 2\pi/L_3$.
In order to trust the PBE, the phonon wavelengths have to be shorter than their mean free paths ($a \ll \ell_D$,
or $L_1 < L_2$).  Otherwise, phonons are not well defined quasiparticles, 
and Boltzmann theory starts to be inapplicable.
The temperature-dependent ratio $L_3/L_2 = k^\ast \ell_D$ is not constrained.  
The local limit (where $k^\ast \ell_D \ll 1$ or $L/\ell_{\rm min}\gg 1$) has phonons seeing
essentially constant thermal gradients, and ordinary local Fourier-law heat transport occurs.  But clean materials
at lower temperatures and small distances from boundaries can be in the opposite regime 
of highly nonlocal (ballistic) transport.

\begin{figure}
\includegraphics[angle=0,width=0.5\textwidth]{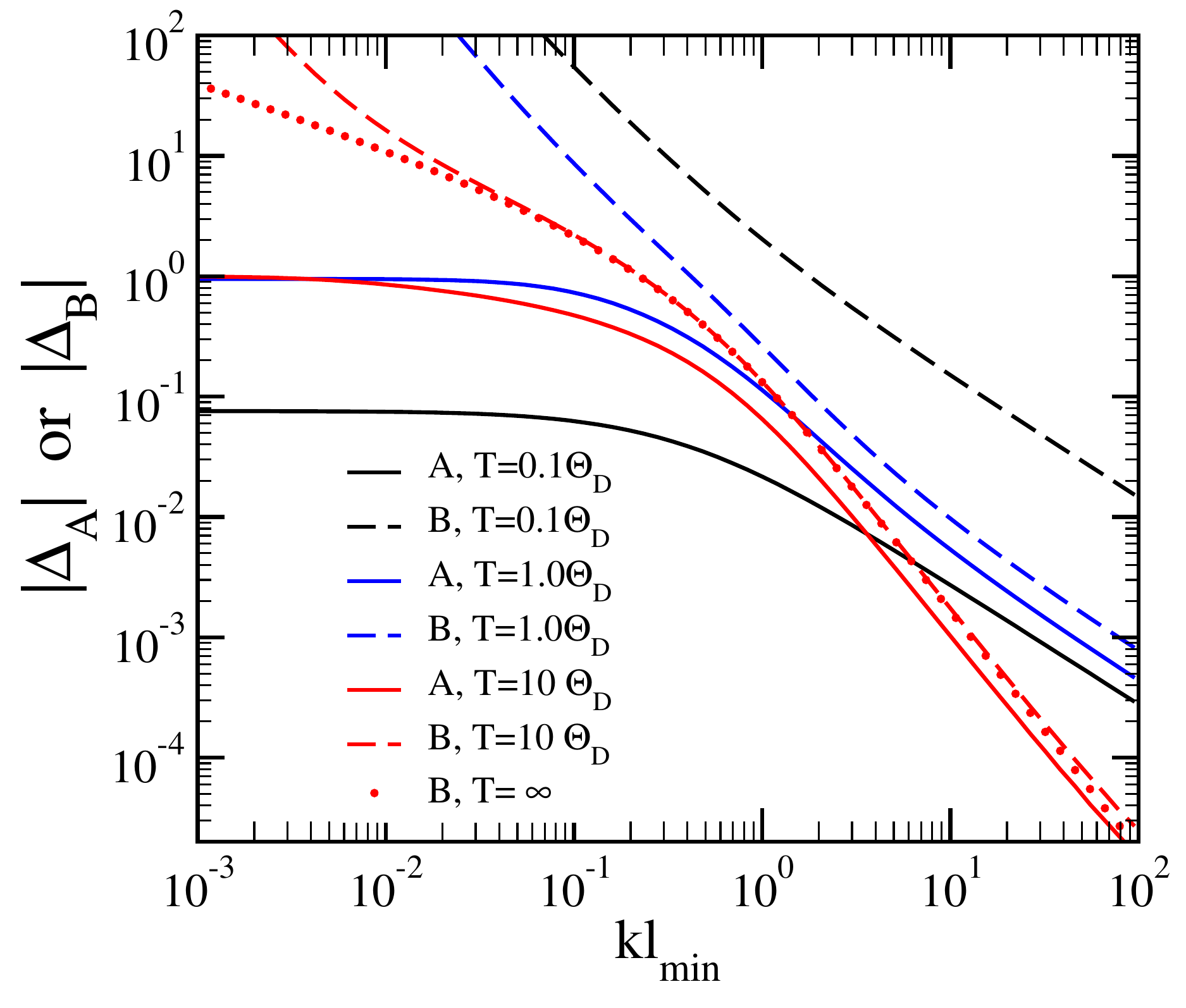}
\caption{\label{fig:mistake} Deviations $\Delta_A$ (Eq. \ref{eq:DA}) and $\Delta_B$ (Eq. \ref{eq:DB}) calculated using the
Debye model (three dimensional, static, $\omega=0$, with $1/\tau_Q\propto q^2$).  Absolute
values are plotted, because $\Delta_B<0$.   }
\end{figure}

One way to compare the two versions is to calculate how much version {\bf A} deviates from
the condition $\langle \hbar\Omega|S_D|\Phi\rangle=0$ required in version {\bf B}, and how
much version {\bf B} deviates from $\langle \hbar\Omega|\Phi\rangle=0$ required in version {\bf A}.
Sensible dimensionless measures are
\begin{equation}
\Delta_A\equiv \frac{1}{V P_0}\langle  \hbar\Omega|S_D|\Phi_A\rangle
\label{eq:DA}
\end{equation}
\begin{equation}
\Delta_B\equiv \frac{1}{V P_0 \tau_D}\langle  \hbar\Omega|\Phi_B\rangle
\label{eq:DB}
\end{equation}
Results are shown in Fig. \ref{fig:mistake}.  The discrepancy $\Delta_A$, which measures what fraction
of input power is lost during scattering (incorrectly treated as inelastic in RTA), is of order 1 in the local (small $k\ell_D$) limit, and gets
small in the highly nonlocal case.  The discrepancy $\Delta_B$, which measures the fraction of the heat (input in one
relaxation time) that is contained incorrectly in deviations from the local equilibrium $n_Q(T(\vec{r}))$, is huge in the
local limit, but diminishes rapidly (except at low $T$) in the nonlocal case.  This pathology in the local limit traces to
a non-analyticity of integrals $S_0$ and $1-S_1$ (defined in the appendix) caused by diverging $\tau_Q\propto 1/q^2$ at small $q$. 
\par
\par
\par
\begin{figure}
\includegraphics[angle=0,width=0.5\textwidth]{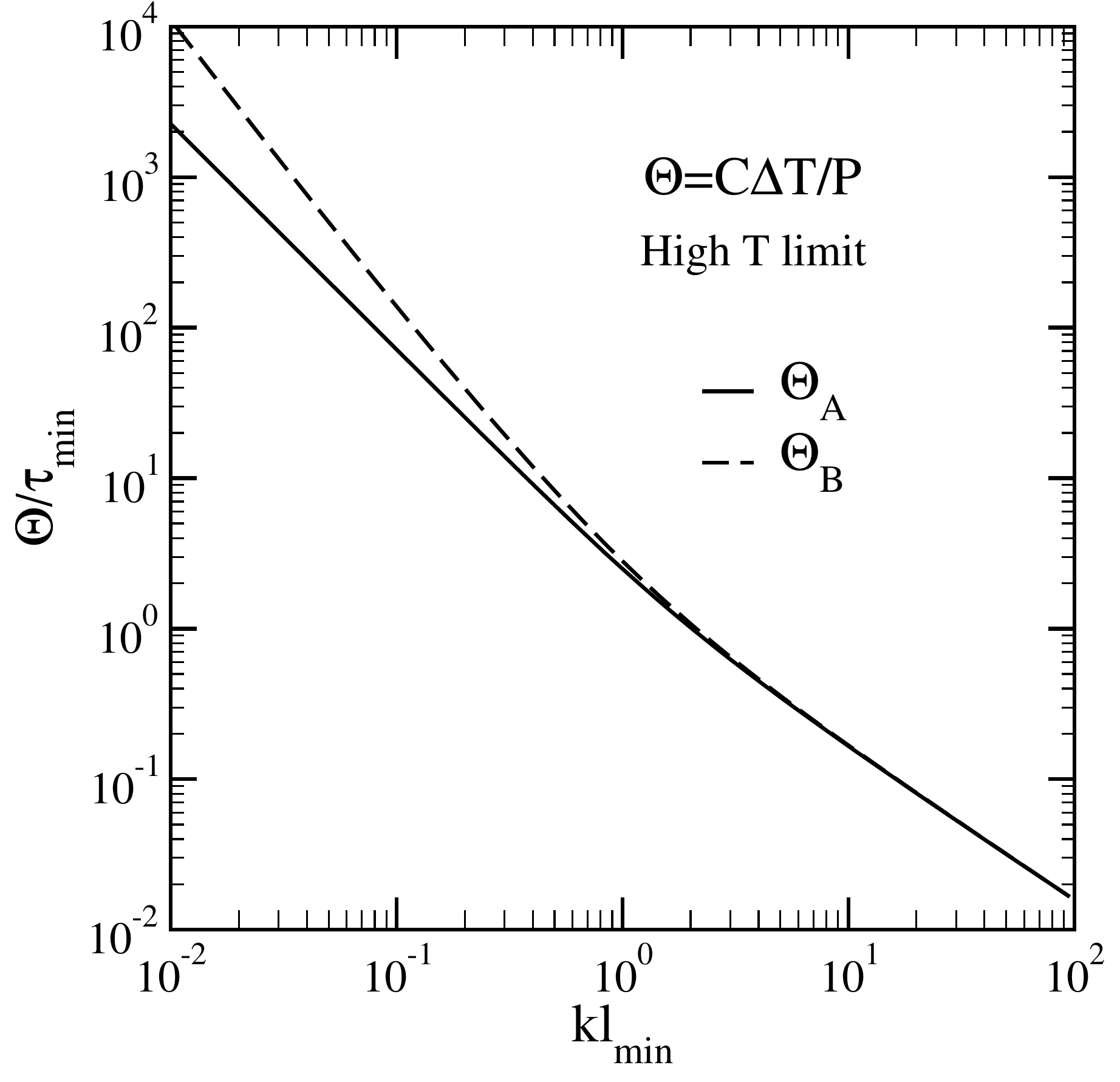}
\caption{\label{fig:42}  Thermal susceptibility $\Theta(k)/\tau_{\rm min}$ 
{\it versus} wavevector $k\ell_{\rm min}$ of the applied heat $P(k)$.
Calculations use the static ($\omega=0$) Debye model with scattering 
$1/\tau_Q =(1/\tau_{\rm min})(q/q_D)^2$. 
$\Theta(k)$ (made dimensionless by dividing by $\tau_{\rm min}$) gives the 
$k$-th Fourier component of the temperature excursion $\Delta T$, per unit applied heat $P/C$.
The calculations are for $T=10 T_D$ where $C\approx 3Nk_B$  in 3 dimensions.  At small $k$, 
$\Theta_A\rightarrow (5\sqrt 2/\pi)(k\ell_{\rm min})^{-3/2}$ and
$\Theta_B\rightarrow (1/k\ell_{\rm min})^2$.  Formulas are discussed in the Appendix.  The
$(k\ell_{\rm min})^{-3/2}$ behavior is unphysical. }
\end{figure}
\par
\par
\begin{figure}
\includegraphics[angle=0,width=0.5\textwidth]{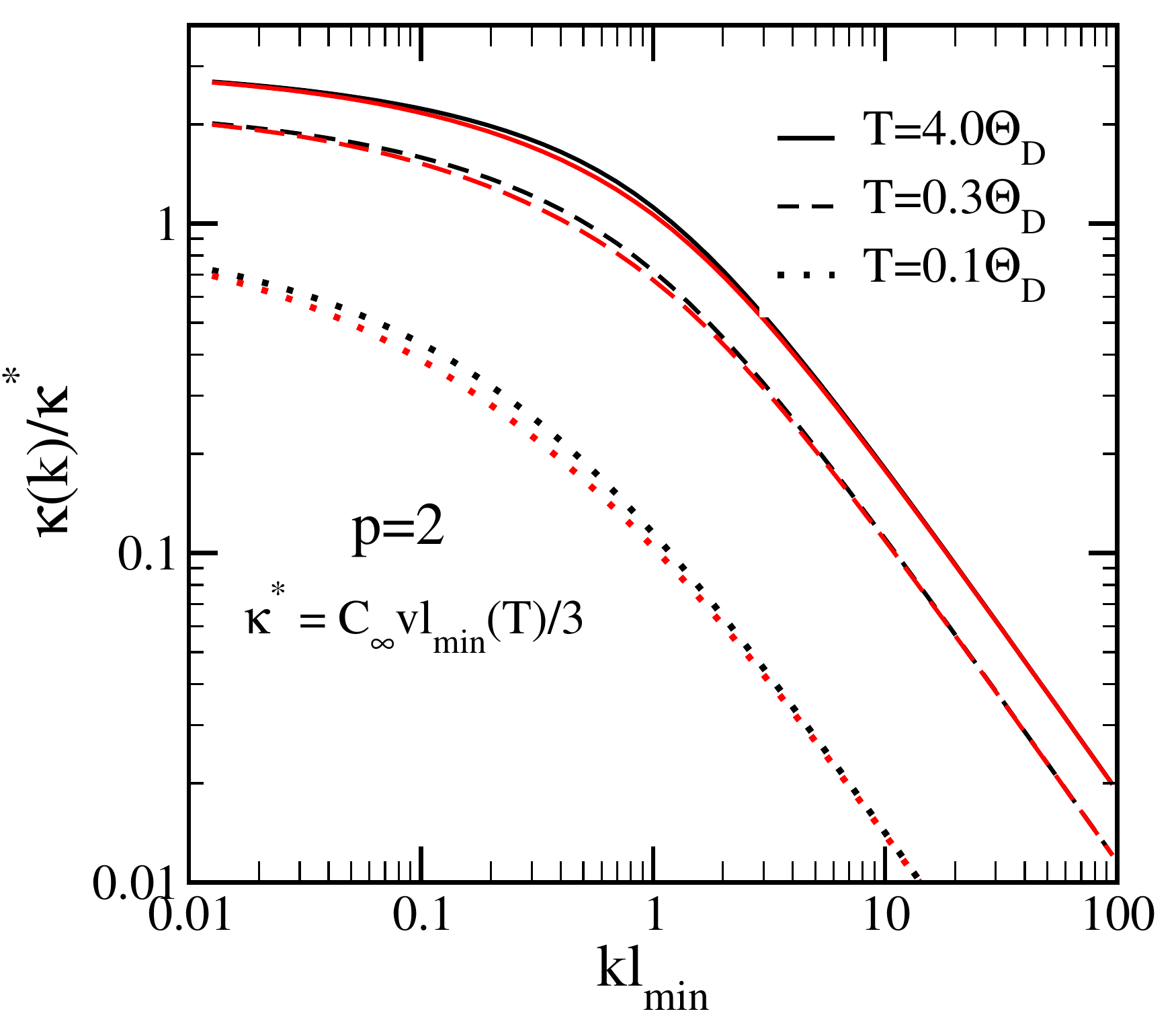}
\caption{\label{fig:kappa}  Thermal conductivity $\kappa(\vec{k})$ {\it versus} $k\ell_{\rm min}$ 
at three temperatures ($T$=4.0,  0.3, and 0.1$ T_D$).  Calculations use 
versions A and B of the RTA in a 3-d Debye model with
$\ell_Q=\ell_{\rm min}(q_D/q)^2$, where $\ell_{\rm min}$ is the minimum phonon mean free path. 
If not normalized to $\kappa^\ast$, the higher $T$ curves would lie below the lower $T$ curves;
$\kappa^\ast$ decreases as $T$ increases, causing the ordering to reverse. 
Black curves are method A and nearby red curves are method B.}
\label{fig:kappa}
\end{figure}
\par

Another way to illustrate the differences between  approaches {\bf A} and {\bf B} is to 
compute thermal susceptibilities $\Theta$.  This is shown in Fig \ref{fig:42}, 
in Debye RTA approximation, with $\Theta$ divided by $\tau_{\rm min}$
to make it dimensionless.    
The two versions {\bf A} and {\bf B} differ significantly at smaller $k$;  version {\bf B} gives correct
physics in this local limit, while version {\bf A} is wrong.  The same pathology of integrals
$S_0$ and $1-S_1$ is responsible, this time for an error in route {\bf A} rather than route {\bf B}.   
The non-analytic pathology appears in $\kappa_{\rm RTA,A}$ but not in $\kappa_{\rm RTA,B}$.  The second
term in Eqs. \ref{eq:RTAkappa} and \ref{eq:RTAkappaB} contains a factor $1/\Theta_A$ (pathological)
and $1/\Theta_B$ (nonpathological).  Fortunately the pathology in $\Theta$ does not show up strongly
in $\kappa$.  This is shown in Fig. \ref{fig:kappa}.

A more physical way of seeing the difference is to examine spatial variations of temperature.  Figure \ref{fig:figgrid}
shows a model with spatial variation having period $L$, allowing Fourier inversion with discrete 
wavevectors $2\pi n/L$.  The Fourier transform of $P(x)$ and the resulting formulas for $\Delta T(x)$ are in
the appendix.    Results are shown in Fig. \ref{fig:Dt}, where the temperature
shift $\Delta T(x)$ is shown for a model with heat input and extraction in regions of size $d=L/8\approx 4\ell_{\rm min}$. 
$\Delta T(x)$ is computed from $\Delta T(k) = \Theta(k)P(k)/C$.   Because of the pathology in $\Theta_A$, the 
results are surprisingly different. 

\par
\begin{figure}
\includegraphics[angle=0,width=0.5\textwidth]{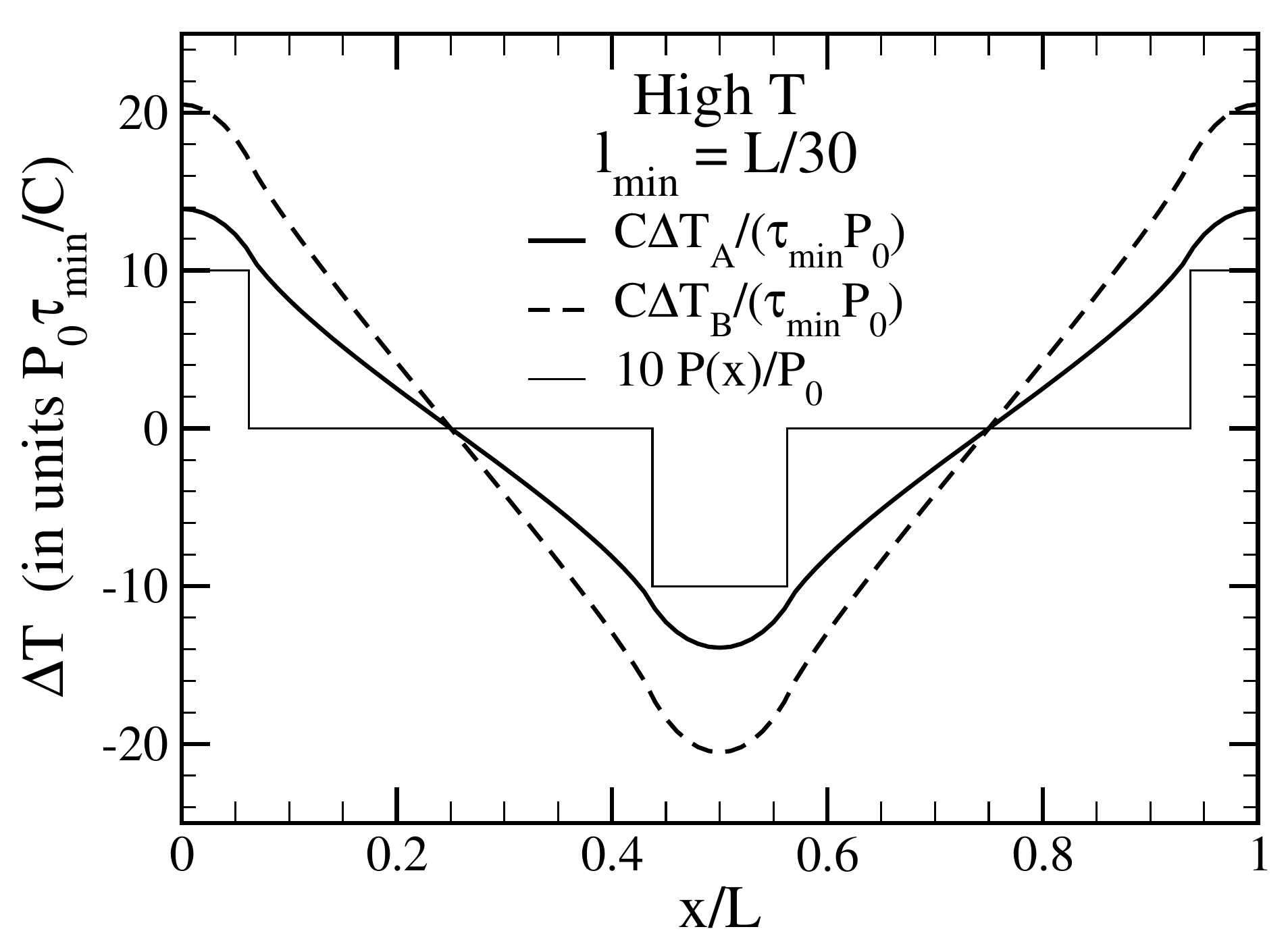}
\caption{\label{fig:Dt} Temperature excursion
$\Delta T$ in RTA {\it versus} position $x$ 
at temperature $T=4.0 T_D$, computed from $\Delta T(k) = \Theta(k)P(k)/C$.  The model ${\bf A}$
results contain an unphysical pathology in $\Theta_A$.
The thin line shows heat input and extraction, $P(x)$.  Only one period of a periodically
repeating system is shown.  The total period is $L=30\ell_ {\rm min}$, where $\ell_ {\rm min}$ is the minimum phonon
mean free path.  Phonons of mode $Q$ have mean free paths $\ell_Q=\ell_ {\rm min}( q_D/ q)^2$.}
\label{fig:Dt}
\end{figure}
\par
\par
\begin{figure}
\includegraphics[angle=0,width=0.5\textwidth]{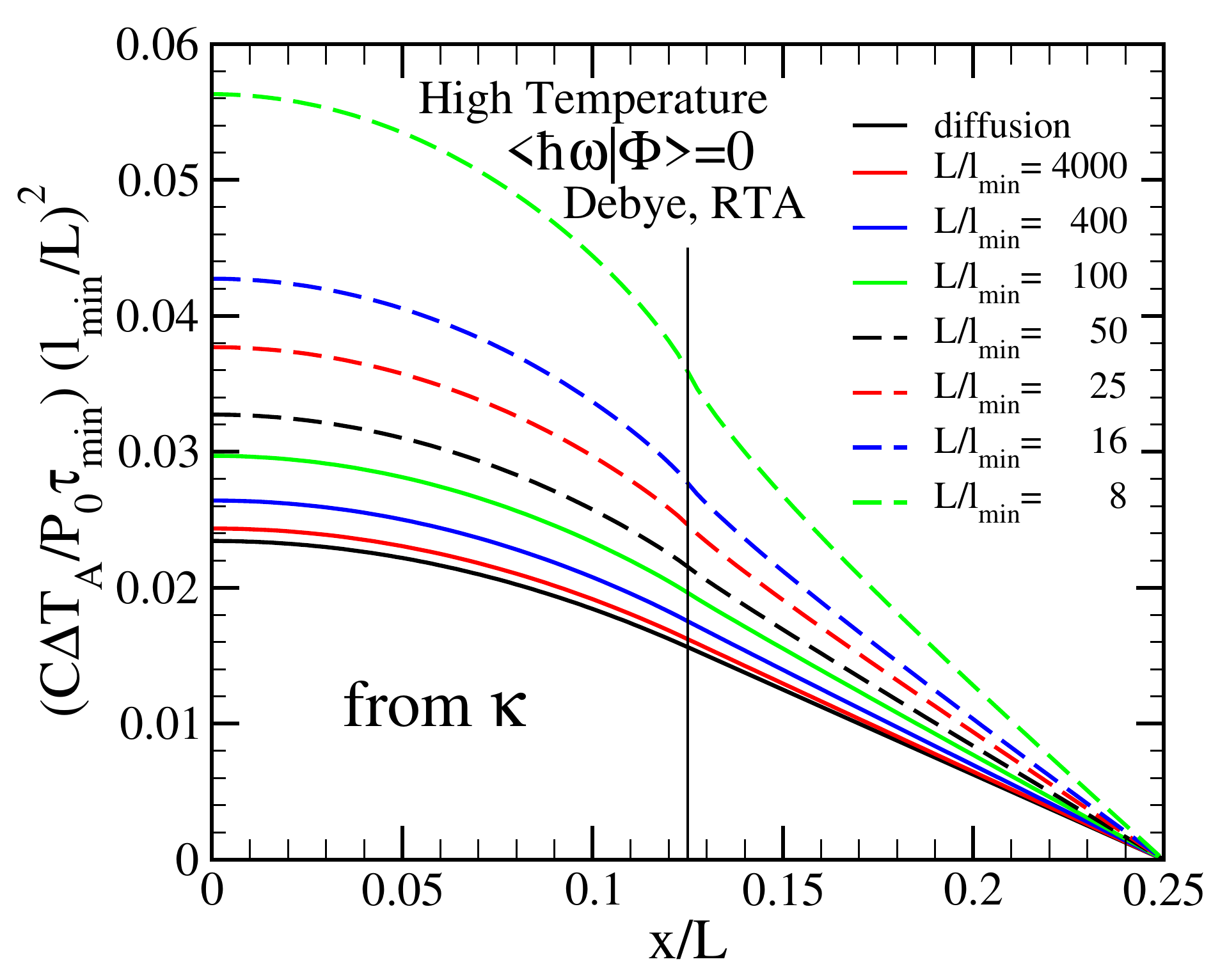}
\caption{\label{fig:DTAAA} Temperature excursion
$\Delta T_A$ in RTA {\it versus} position $x$, from model A, using $\Delta T(k) = P(k)/k^2\kappa_L(k)$.
The temperature is $T=10.0 T_D$,  $\ell_Q=\ell_{\rm min}(q_D/q)^2$, and $L/\ell_{\rm min}$ takes a range of values.
The lowest values (or highest effective conductivity) are in the large $L/\ell_{\rm min}$, or diffusive, limit.
Heat $P(x)$ is added in the region $x<d/2=L/8$; $P(x)=0$ in the region $x>d/2$.  The thin
vertical line marks $d/2=L/8$.  Only one quarter period of a periodically repeating system is shown.}
\end{figure}
\par

\par
\begin{figure}
\includegraphics[angle=0,width=0.5\textwidth]{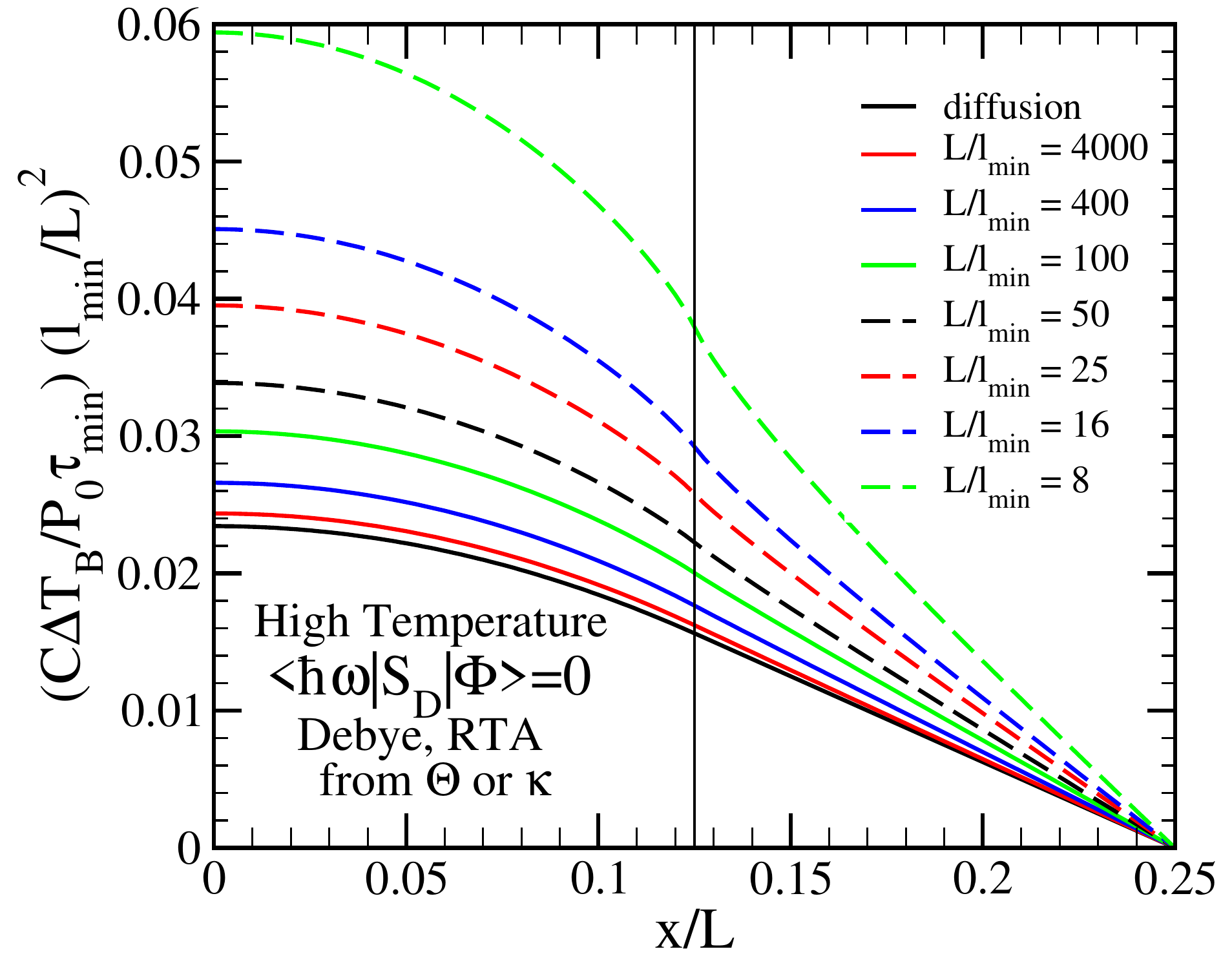}
\caption{\label{fig:DTBB} Temperature excursion
$\Delta T_B$ in RTA {\it versus} position $x$, from model B, using either $\Delta T(k) = P(k)/k^2\kappa_L(k)$,
or $\Delta T(k)=\Theta_B(k) P(k)/C$ (the formulas are identical in model B).
The parameters are the same as used in Fig. \ref{fig:DTAAA}.}
\end{figure}
\par

The difference is much smaller when   
$\Delta T(x)$ is computed from $\Delta T(k) = P(k)/k^2\kappa_L(k)$, where the longitudinal part of $\kappa$ is
$\kappa_L=\hat{k}\cdot\kappa\cdot\hat{k}$.  Figures \ref{fig:DTAAA} and \ref{fig:DTBB} show such calculations, in the
high $T$ classical limit, for a range of $L/\ell_{\rm min}$.  Both routes {\bf A} and {\bf B} converge correctly to the diffusive limit
for large values of $L/\ell_{\rm min}$, and their predictions for $\Delta T(x)$ are quite similar, deviating a bit from each
other in the non-local case of smaller $L/\ell_{\rm min}$.

\section{conclusions}

We have considered the simplest sensible model for the heat input term $(\partial N_Q/\partial t)_{\rm ext}$
needed to get nonlocal features in phonon Boltzmann theory.  The concept of thermal susceptibility
$\Theta(\vec{k},\omega)=C(T)d\Delta T(\vec{k},\omega)/dP(\vec{k},\omega)$ is a natural consequence of
dealing with external driving, but depends on how the driving is modeled.  Exact Boltzmann theory is unambiguous
about the definition of local temperature $T(\vec{r},t)$ or $T(\vec{k},\omega)$.  However, when treated in 
RTA, an ambiguity seems inevitable.  If temperature is constrained by forcing relaxation to the local
Bose-Einstein distribution $n(\hbar\omega_Q/k_B T(\cdot))$, as demanded by the exact theory, the RTA
version is less internally consistent than desireable.  If instead, temperature is constrained by forcing
the energy change caused by collisions, $(dE/dt)_{ \rm scatt}$, to vanish, the result is more internally
consistent even though at odds with the exact procedure.  The predicted nonlocal variation of $\Delta T$
is reasonably similar for the two definitions.

\section{acknowledgements}

 We thank A. G. Abanov, R. Chamberlin, Chengyun Hua, J. P. Nery, K. K. Likharev, and M. K. Liu for helpful conversations.
This work was supported in part by DOE grant No. DE-FG02-08ER46550.

\section{Appendix: Details}

\subsection{$\Delta T(x)$ in the diffusive regime}

If the applied heating $P$ is independent of time, then the steady state solution has $\partial U/\partial t =0$,
and Eq. \ref{eq:Econsv} says $P=\vec{\nabla}\cdot\vec{j}$.  In the diffusive regime, the Fourier law is
$\vec{j}=-\kappa\cdot\vec{\nabla}T$.  For heating $P=P(x)$ varying only in one dimension, the
temperature then obeys $d^2 T/dx^2 = -P(x)/\kappa$.  Solving this for the case illustrated in Fig. \ref{fig:figgrid}
requires using the fact that for $d/2<x<L/4-d/2$, where $P=0$, the current is constant, $Pd/2$,
and the temperature gradient is $-Pd/2\kappa$.  The temperature $T=T_0 + \Delta T(x)$ has
$\Delta T(d/4)=0$ at the midpoint between heat injection and removal.  Then the answer, which is plotted
in Figs. \ref{fig:figgrid}, \ref{fig:DTAAA}, and \ref{fig:DTBB} is
\begin{eqnarray}
\Delta T(x)&=&\frac{Pd}{2\kappa}\left(\frac{L}{4}-x\right), \ \ \ \frac{d}{2}<x<\frac{L}{2}-\frac{d}{2} \nonumber \\
&=& \frac{P}{2\kappa}\left(\frac{d}{4}(L-d)-x^2\right), \ \ \ |x|<\frac{d}{2}
\label{eq:Dtdiff}
\end{eqnarray}

\subsection{Effective thermal Conductivity}

For the geometry of Fig. \ref{fig:figgrid}, there are several possible ways to define an effective thermal conductivity.
The heat current is known from the total steady state heat input.  Energy $Pd$ enters per unit time and is carried
away, but only half is carried as $j_x > 0$ to the right ($x>d/2$) and half to the left ($x<-d/2$).  
So $j_x$ between heat injection and removal is $\pm Pd/2$.  What temperature
gradient is to be taken?  In the region near $x=L/4$, in the diffusive regime, many mean free paths distant from the heaters, 
the gradient is $-Pd/2\kappa$, and the ratio $j/(-dT/dx)$ is just the bulk $\kappa$.  But in a nanoscale experiment, 
perhaps the more relevant measure is to take the average gradient to be the total temperature difference
$\Delta T(x=0)-\Delta T(x=L/2) = 2\Delta T(0)$ between heater and cooler and divide by the distance $L/2$.
This is a smaller gradient and therefore corresponds to a larger effective conductivity $\kappa/(1-d/L)$.
Or perhaps good thermometry can deliver the average temperature over the region ($-d/2<x<d/2$) of
the heater.  Again dividing by distance $L/2$ gives an effective conductivity $\kappa/(1-7d/6L)$,
a bit higher.  Then again, maybe twice this average temperature peak height is to be divided by the distance
$L/2-d$ of heat flow between heater and cooler.  This gives an effective conductivity
$\kappa\times (1-2d/L)/(1-7d/6L)$, smaller than the bulk value.  Other possibilities can be imagined.
These values of $\kappa_{\rm eff}$ are for purely diffusive transport.  Similar ambiguities, with different answers, occur for 
``quasiballistic'' transport when mean free paths are no longer negligibly small.

At the level of fundamental theory, a nonlocal $\kappa(\vec{k},\omega)$ should describe everything,
although the theory for $\kappa(\vec{k},\omega)$ can change with different models for the power input
$P(\vec{r},t)$.  This discussion suggests that for nanoscale heat problems, $\kappa$ is not the 
clearest choice of analytic tools.  The temperature rise per unit power input, $\Delta T(\vec{k},\omega)/P(\vec{k},\omega)$,
which has been labeled in Eq. \ref{eq:defTheta} as $\Theta(\vec{k},\omega)/C(T)$, with $\Theta$ the thermal susceptibility, 
is a more direct measure of the interesting properties of the nanosystem.

\subsection{Debye model integrals in RTA}

For calculations using RTA, we need to evaluate various integrals of the form
\begin{equation}
F_{mn}(k,T)=\frac{1}{3N}\sum_Q\left(\frac{z}{\sinh z}\right)^2 
\frac{(v_{Qx}/v)^m  (\tau_D/\tau_Q)^n}{(\tau_D/\tau_Q)+ikv_{Qx}\tau_D}
\label{eq:Fmn}
\end{equation}
Factors of minimum relaxation time $\tau_D$ and sound velocity $v$ are introduced to make $F_{mn}$ dimensionless. 
Use was made of the harmonic specific heat formula
\begin{equation}
C(T)=\frac{k_B}{V}\sum_Q \left( \frac{z}{\sinh z} \right)^2 
\label{eq:spht}
\end{equation}
where $z=\hbar\omega_Q/2k_B T$.  In the high $T$ (classical) limit, with one atom per primitive cell,
$C\rightarrow C_\infty= 3Nk_B /V$, since the sum over modes $\sum_Q=3N$ in three dimensions with
one atom per cell.   The assumptions have
been introduced that $P$ is time-independent (so $\omega=0$), and varies only in the $x$ direction.
Only the longitudinal component $\kappa_{xx}(k)$ is examined, where $\vec{k}=k\hat{x}$. The RTA formulas
Eqs. \ref{eq:RTAkappa}, \ref{eq:ThetaA}, \ref{eq:ThetaB}, and \ref{eq:RTAkappaB} can be written as
\begin{equation}
\kappa_A(k)=3\kappa^\ast \left(F_{20}-\frac{F_{10}^2}{F_{00}}\right) 
\label{eq:kA}
\end{equation}
\begin{equation}
\kappa_B(k)=3\kappa^\ast \left(F_{20}-\frac{F_{10} F_{11}}{F_{01}}\right)
\label{eq:kB}
\end{equation}
\begin{equation}
\Theta_A = \frac{F_{00}}{ikv F_{10}} \ {\rm and} \ \Theta_B = \frac{F_{01}}{ikv F_{11}}
\label{eq:ThetaAB}
\end{equation}
where the thermal conductivity scale $\kappa^\ast=C_\infty v^2 \tau_D/3$ is introduced.
In the Debye model, with $1/\tau_Q=(1/\tau_D)(q/q_D)^p$, $F_{mn}$ becomes
\begin{equation}
F_{mn,D}=\frac{1}{N}\sum_{\vec{q}} \left(\frac{z}{\sinh z}\right)^2 \frac{\mu^m (q/q_D)^{np}}
{(q/q_D)^p +ik\ell_D \mu},
\label{eq:FmnD}
\end{equation}
where $\ell_D=v\tau_D$ is the Debye smallest mean free path.
This uses $\sum_Q = 3\sum_{\vec{q}}$ because there are $3N$ modes labeled by $Q$ and $N$ wavevectors
labeled by $\vec{q}$.  The angular integral uses
$\mu=\cos\theta$, $\theta$ being the angle between $\vec{v}_q$ or $\vec{q}$ and $\hat{x}$.
Now let $u=q/q_D$.  The variable $z$ in the specific heat is $\hbar\omega_q/2k_B T
=u T_D /2T$.  The integral can be written
\begin{equation}
\frac{1}{N}\sum_{\vec{q}} = \frac{3}{2}\int_0^1 du u^2 \int_{-1}^1 d\mu.
\label{eq:sumq}
\end{equation}
There are three angular integrals ($m = 0,1,2$),
\begin{equation}
A_m = \frac{1}{2}\int_{-1}^{1} d\mu \frac{\mu^m}{u^p +ik\ell_D \mu} 
\label{eq:Am}
\end{equation}
\begin{equation}
A_0= \frac{1}{k\ell_D} \tan^{-1}\left(\frac{k\ell_D}{u^p}\right)
\label{eq:mu0}
\end{equation}
\begin{equation}
A_1= \frac{1}{ik\ell_D} (1-u^p A_0)
\label{eq:mu1}
\end{equation}
\begin{equation}
A_2 = \frac{u^p}{(k\ell_D)^2}(1-u^p A_0)
\label{eq:mu2}
\end{equation}
Equation \ref{eq:FmnD} is then 
\begin{equation}
F_{mn,D}(k)=3\int_0^1 du \left( \frac{z}{\sinh z} \right)^2 u^{2+np} A_m(u,k)
\label{eq:FmnD1}
\end{equation}
The formulas for $\Theta$ in Eqs. \ref{eq:ThetaA}, \ref{eq:ThetaB}, and \ref{eq:ThetaAB} can now be written as
\begin{equation}
\Theta_{\rm RTA,A}(k)/\tau_{D}=\frac{S_0(k)}{R_0-S_1(k)}
\label{eq:chiA}
\end{equation}
\begin{equation}
\Theta_{\rm RTA,B}(k)/\tau_{D}=\frac{S_1(k)}{R_1-S_2(k)}
\label{eq:chiB}
\end{equation}
where
\begin{equation}
R_n = 3\int_0^{ q_D} \frac{d q  q^2}{ q_D^3} \left(\frac{z}{\sinh z}\right)^2 \left(\frac{ q}{ q_D}\right)^{np}
\label{eq:Rn}
\end{equation}
\begin{equation}
S_n(k)= 3\int_0^{ q_D} \frac{d q  q^2}{ q_D^3} \left(\frac{z}{\sinh z}\right)^2 
\frac{\tan^{-1}\left[k\ell_D \left(\frac{ q_D}{ q}\right)^p \right]}{k\ell_D} \left(\frac{ q}{ q_D}\right)^{np}.
\label{eq:Sn}
\end{equation}
Here the $p$-dependent functions $S_n$ and $R_n$ depend on $k\ell_D(T)$ and $T/ T_D$, where the
Debye temperature is $\hbar v  q_D/k_B$.   

In the high $T$ limit where $(z/\sinh z)^2 \rightarrow 1$,  $R_n = 3/(3+np)$. 
The $S_n$ integrals can also be done analytically at high $T$ for $p=2$.  The full formulas  are messy and give
little insight, but the small $k$ limits can be extracted and used to show that $\Theta_A \rightarrow
5\sqrt{2}/\pi (k\ell_D)^{3/2}$, and $\Theta_B \rightarrow 1/(k\ell_D)^2$.
Both agree well with numerics in Fig. \ref{fig:42}.  The non-analytic behavior of $\Theta_A$ at small $k$
is caused by the peculiar behavior of the arctangent function in Eq. \ref{eq:Sn}, when
$p>0$ and both $k$ and $q$ are small.  The extra powers of $(q/q_D)^{np}$ for $n=2$ suppress
the non-analyticity, but for $n$=1 it causes $\Theta_A$ to be badly behaved, and destroys diffusive behavior
in $\Theta_A$.  The small $k$ diffusive behavior is given correctly by $\Theta_B\rightarrow 1/(k\ell_D)^2$,

\subsection{Non-diffusive $\Delta T(x)$ by Fourier inversion}

The spatial behavior of $\Delta T$ is shown in
Figs. \ref{fig:Dt}, \ref{fig:DTAAA}, and \ref{fig:DTBB}, for the heating configuration
of Fig. \ref{fig:figgrid}.  The formula is
\begin{eqnarray}
\Delta T_\Theta (\vec{r}) &=&\sum_{\vec{k}} \frac{CP(\vec{k})}{\Theta(\vec{k})} \ e^{i\vec{k}\cdot\vec{r}};  \nonumber \\
\Delta T_\kappa(\vec{r})&=&\sum_{\vec{k}} \frac{P(\vec{k})}{\vec{k}\cdot\kappa(\vec{k})\cdot\vec{k}} \ e^{i\vec{k}\cdot\vec{r}}
\label{eq:FT}
\end{eqnarray}
The two versions are the same for route {\bf B}, but in route {\bf A}, only the second ($\Delta T_\kappa$)
should be used, to avoid the incorrect small $k$ behavior of $\Theta_A$ in RTA.
In the one-dimensional heating arrangement, the Fourier vector $\vec{k}=(2\pi n/L,0,0)$
has only an $\hat{x}$ component.  The system is spatially homogeneous in the $\hat{y}$ and $\hat{z}$
directions ($k_y=k_z=0$).  The periodicity $L$ in the $\hat{x}$ direction means that
$k_x$ is quantized in units $2\pi/L$.  $P(x)$ and $\Delta T(x)$ are both even in $x$, so
that the $-k$ and $+k$ parts of the $k$-sums in Eq. \ref{eq:FT} can be combined, and
$\exp(ikx)+\exp(-ikx)$ replaced by $2\cos(kx)$.  Finally, $P(x)$ and $\Delta T(x)$ are both
antisymmetric around $x=L/4$.  This causes $P(k)$ and $\Delta T(k)$ to vanish when
the integer $n$ is even.  The equation for $P(k)$ is found from
\begin{equation}
P(k)=\frac{1}{L}\int_0^L dx P(x)e^{-ikx}
\label{eq:Pk}
\end{equation}
where the input power is 
\begin{equation}
P(x)=P_0 [\theta(d/2-|x|)-\theta(d/2-|x-L/2|)],
\label{eq:Pxreal}
\end{equation}
and $\theta(x)$ is the Heaviside unit step function.  Then $P(k)$ is
\begin{equation}
P(k)=P_0 \frac{2\sin(kd/2)}{kL/2} \ {\rm for} \ k=\frac{2\pi n}{L} \ {\rm with} \ n \ {\rm an \ odd \ integer}.
\label{eq:Pkreal}
\end{equation}
With these equations, the Fourier inversion can be done.

\bibliography{temperature}

\end{document}